\documentclass[prb,twocolumn,showpacs]{revtex4}
\usepackage{graphicx}
\usepackage{dcolumn}
\usepackage{bm}
\begin{document}
\title{Magnetic heat transport in {\it R}$_2$CuO$_4$ ({\it R} = La, Pr, Nd, Sm, Eu, and Gd)}


\author{K. Berggold}
\author{T. Lorenz}
\author{J. Baier}
\author{M. Kriener}
\author{D. Senff}
\author{H. Roth}
\author{A. Severing}
\author{H. Hartmann}
\author{A. Freimuth}
\affiliation{$ $II. Physikalisches Institut, Universit\"{a}t zu K\"{o}ln,
Z\"{u}lpicher Str. 77, 50937 K\"{o}ln, Germany}
\author{S. Barilo}
\affiliation{Institute of Solid State $\&$ Semiconductor Physics,
Belarussian Academy of Sciences, Minsk 220072, Belarus}
\author{F. Nakamura}
\affiliation{Department of Quantum Matter, ADSM, Hiroshima University,
Higashi-Hiroshima, 739-8526 Japan}

\date{\today}
\begin{abstract}
We have studied the thermal conductivity $\kappa$ on single
crystalline samples of the antiferromagnetic monolayer cuprates
{\it R}$_2$CuO$_4$ with {\it R}~=~La, Pr, Nd, Sm, Eu, and Gd. For
a heat current within the CuO$_2$ planes, i.\,e. for
$\kappa_{ab}$ we find high-temperature anomalies around 250\,K in
all samples. In contrast, the thermal conductivity $\kappa_c$
perpendicular to the CuO$_2$ planes, which we measured for $R$ =
La, Pr, and Gd, shows a conventional temperature dependence as
expected for a purely phononic thermal conductivity. This
qualitative anisotropy of $\kappa_i$ and the anomalous temperature
dependence of $\kappa_{ab}$ give evidence for a significant
magnetic contribution $\kappa_{\rm mag}$ to the heat transport
within the CuO$_2$ planes. Our results suggest, that a large
magnetic contribution to the heat current is a common feature of
single-layer cuprates. We find that $\kappa_{\rm mag}$ is hardly
affected by structural instabilities, whereas already weak charge
carrier doping causes a strong suppression of $\kappa_{\rm mag}$.
\end{abstract}

\maketitle
\section{introduction}
Various low-dimensional spin systems show an unusual thermal
conductivity $\kappa$ with a double-peak structure as a function
of temperature. There is growing evidence that this anomalous
behavior arises from magnetic excitations contributing to the
heat transport. The most clear experimental evidence is found in
the spin-ladder compounds, where a double-peak structure with a
huge high-temperature maximum of $\kappa$ is present for a heat
current parallel to the ladder direction, but absent for a heat
current perpendicular to the
ladders.\cite{sologubenko00a,hess01a} For one-dimensional
spin-chain systems the experimental results are less clear. For
the spin-chain compounds SrCuO$_2$ and Sr$_2$CuO$_3$ Sologubenko
{\it et al.} find a sizeable extra contribution to $\kappa$ along
the chain direction, which is missing in the other
directions.\cite{sologubenko01a} In the spin-Peierls compound
CuGeO$_3$ $\kappa$ along the chain direction has two
low-temperature maxima and one of them was attributed to a
magnetic contribution in Ref.~\onlinecite{ando98a}. However, this
interpretation is questionable, because a similar double peak is
also present in $\kappa$ perpendicular to the chain
direction.\cite{hofmann02a} For the Haldane-chain ($S=1$) System
AgVP$_2$S$_6$, a magnetic contribution seems to play a role
too,\cite{sologubenko03a} but the absolute values are much
smaller than in the $S=1/2$ systems. The results in 1D systems
raise the question whether a sizeable heat current due to
magnetic excitations is also present in two-dimensional magnets.
This was discussed for the low-temperature thermal conductivity
of K$_2$V$_3$O$_8$ and Nd$_2$CuO$_4$.\cite{sales02a,jin03a,li05a}
The latter is one of the insulating parent compounds of
high-temperature superconductors containing CuO$_2$ planes, which
represent the perhaps most studied two-dimensional
antiferromagnets so far.\cite{kastner98a} Whereas the studies on
Nd$_2$CuO$_4$ (Refs.~\onlinecite{jin03a,li05a}) mainly concern
the magnetism of the Nd$^{3+}$ moments, the influence of the
Cu$^{2+}$ moments is present at higher temperature. In the
layered perovskite La$_{2}$CuO$_{4}$, the thermal conductivity
$\kappa_{ab}$ for a heat current along the CuO$_2$ planes
exhibits a pronounced double-peak structure with a
low-temperature maximum around 25\,K and a second one around
250\,K. In contrast, the thermal conductivity $\kappa_{c}$
perpendicular to the CuO$_2$ planes has only one low-temperature
peak.\cite{nakamura91a,yan03a,sun03a,hess03a} These findings have
been interpreted in terms of an additional heat transport
parallel to the CuO$_2$ planes due to magnetic excitations.
However, a double-peak structure can also be explained by phonons
only. Any additional scattering mechanism which acts in a narrow
temperature range suppresses $\kappa$ in that temperature window
and as a result $\kappa$ may exhibit two peaks. For example, in
SrCu$_2$(BO$_3$)$_2$ a double-peak structure is caused by
resonant scattering of acoustic phonons by magnetic
excitations.\cite{hofmann01a}  Such a mechanism does not apply
for La$_{2}$CuO$_{4}$. However, La$_{2}$CuO$_{4}$ has a
structural instability with low-lying optical phonon branches,
which could also serve as scatterers for the acoustic phonons.
Such an explanation has been proposed by Cohn \textit{et al.} for
the heat transport data of YBaCuO$_{6+\delta}$, which show a
similar temperature dependence of $\kappa_{ab}$  as
La$_2$CuO$_{4}$.\cite{cohn95a}

Recently, we have measured $\kappa_{ab}$ of Sr$_2$CuO$_2$Cl$_2$
in order to investigate the possibility that resonant phonon
scattering due to the structural instability may cause the
double-peak structure in $\kappa_{ab}$ of La$_{2}$CuO$_{4}$.
Sr$_2$CuO$_2$Cl$_2$ and La$_{2}$CuO$_{4}$ are almost
isostructural, but Sr$_2$CuO$_2$Cl$_2$ has no structural
instability. Nevertheless we also found a second high-temperature
maximum in $\kappa_{ab}$ and took this as evidence for the second
peak of the in-plane heat conductivity being caused by magnetic
excitations.\cite{hofmann03a} Basing on this observation we
expect that a pronounced magnetic contribution to the thermal
conductivity is a common feature of the layered cuprates.
However, the high-temperature peak of $\kappa_{ab}$ of
La$_{2}$CuO$_{4}$ is significantly larger and its low-temperature
peak is much smaller than the corresponding peaks observed in
Sr$_2$CuO$_2$Cl$_2$. These quantitative differences could arise
from the absence or presence of a structural instability and/or
weak charge carrier doping in the different samples. In order to
investigate whether the high-temperature peak of $\kappa_{ab}$ is
indeed an intrinsic feature of the antiferromagnetic CuO$_2$
planes and to get more insight about the influence of structural
changes, we have studied the thermal conductivity on single
crystals of $R_2$CuO$_4$ with different rare earths $R$, which
realize different structures. Some of these compounds even show a
structural transition as a function of
temperature.\cite{johnston97a,braden94b,braden94a,vigoureux97b} Up
to now only the in-plane thermal conductivity for $R={\rm Pr}$ and
Nd has been
studied,\cite{sologubenko99a,inyushkin94a,jin03a,li05a} and the
studies of $\kappa_{ab}$ and $\kappa_{c}$ of La$_2$CuO$_4$ have
concentrated on the influence of
doping.\cite{nakamura91a,yan03a,sun03a,hess03a}

Here, we present measurements of $\kappa_{ab}$ for $R = {\rm
Pr}$, Nd, Sm, Eu, and Gd and of $\kappa_{c}$ for $R = {\rm Pr}$
and Gd. Our study clearly shows that the anomalous double-peak
structure of $\kappa_{ab}$ is present for all $R_2$CuO$_4$ and
confirms that a large magnetic contribution to the heat transport
is an intrinsic property of the CuO$_2$ planes. In addition, we
find that this magnetic contribution is hardly affected by a
structural instability while the phononic contribution is strongly
suppressed.

\section{Magnetic and structural properties of $R_{2}$$\mbox{CuO}_{4}$}
\label{seccomp}

A common feature of La$_{2}$CuO$_{4}$ and $R_{2}$CuO$_{4}$, with
$R = {\rm Pr}$, Nd, Sm, Eu, and Gd is the layered structure with
planes consisting of a CuO$_2$ square lattice. These planes are a
good realization of a two-dimensional antiferromagnetic
$S=\frac{1}{2}$ Heisenberg square lattice. Bi-magnon Raman
scattering\cite{cooper90a} yields exchange constants $J$ between
$\approx 1200-1400$\,K for $T=300$\,K given in
Table\,\ref{tabelle}.\cite{roomtemp} Finite inter-plane couplings
$J_\perp$ lead to three-dimensional antiferromagnetic ordering
with N\'{e}el temperatures of $T_{\rm N}\approx 250 \dots 320$\,K
(see Table\,\ref{tabelle}). In general, the ordering temperature
is determined by the ratio of the inter- and the intra-plane
coupling. However, crystal quality and the oxygen stoichiometry
strongly influence $T_{\rm N}$, too. For example, very small
amounts of excess oxygen drastically suppress $T_{\rm N}$ of
La$_2$CuO$_{4+\delta}$.\cite{chen91a} Apart from the magnetic Cu
subsystem, the compounds with magnetic rare earth ions $\it R$
contain another magnetic subsystem.  In all these compounds the
behavior of the Cu subsystem is very similar. However, the
details of the magnetic structure are determined by the
competition between the different couplings (Cu-Cu, $\it R$-$\it
R$, $R$-Cu). \cite{sachidanandam97a,petitgrand99a}

\begin{table*}
\caption{\scriptsize Exchange constants $J$ at $300$\,K (Ref.
\onlinecite{cooper90a}), Debye temperatures $\Theta_D$, sound
velocities $v_s$, sample sizes, and  N\'{e}el temperatures $T_{\rm
N}$ of $R_2$CuO$_4$ (see text and the respective references). $P$,
$D$, $U$, and $u$ are fit parameters for the fits of the phononic
contribution of $\kappa $. $P$ and $D$ describe the scattering on
point defects and planar defects, respectively, whereas $U$ and
$u$ model Umklapp scattering (for details see
Ref.~\onlinecite{hofmann03a}). If available, the values for
$\Theta_D$ and $v_s$ are taken from literature, otherwise similar
values have been used for the fits.}
 \label{tabelle}
  \scriptsize
\vskip1mm
\parbox{16.5cm}{
\begin{ruledtabular}
\begin{tabular}{llccccccccc}
 &  &$J$ & $\Theta_D$&  $v_s$ &  $a \times b \times
c$&$T_{\rm N}$  &$P$ & $D $ & $U $ & $u$ \\
 & & (K)  & (K)& (m/s) &(mm$^3$)&(K) & (10$^{-43}$s$^3$)&$(10^{-18}$s)
&$(10^{-31}$s$^2/$K$)$& \\
\hline
La$_{2}$CuO$_{4}$ $ab$ &(Ref.\,\onlinecite{nakamura91a})&$1465$&$385$ (Ref. \onlinecite{junod89a}) &$5200$ (Ref. \onlinecite{suzuki00a})&&$316$&$21$&$26$&$22$&$4.4$\\
La$_{2}$CuO$_{4}$ $ab$ &(Ref.\,\onlinecite{yan03a})&$$&$$&&&$323$&$ $ $1.5$&$11.8$&$23$&$5.3$\\
La$_{2}$CuO$_{4}$ $ab$ &(Ref.\,\onlinecite{yan03a})&$$&$$&&&$313$&$ $ $23.8$&$4.6$&$14.1$&$7$\\
La$_{2}$CuO$_{4}$ $ab$ &(Ref.\,\onlinecite{sun03a})&$$&$ $&&&$308$&$$$25.8$&$26.2$&$15.3$&$4.4$\\
La$_{2}$CuO$_{4}$ $c$ &(Ref.\,\onlinecite{nakamura91a}) &&&&&$316$&$ $ $15.4$&$14.6$&$17.3$&$6.4$\\
La$_{2}$CuO$_{4}$ $c$ &(Ref.\,\onlinecite{yan03a})
&&&&&$325$&$1.9$&$15.1$&$18.8$&$5.5$\\
La$_{2}$CuO$_{4+\delta}$ & &&$$&&$0.6\times3\times2.5$&$245$&
&$$&$$&$$\\[\medskipamount]
Pr$_{2}$CuO$_{4}$ $ab$ &&$1243$&$361$
 (Ref. \onlinecite{hundley89a})&$6000$
 (Ref. \onlinecite{fil96a})&$1.7 \times
1.6\times1.4$&$250$&$7.1$&$1$&$8.5$&$5.8$\\
Pr$_{2}$CuO$_{4}$ $c$ & &&&&&&$10.3$&$2.0$&$13.0$&$5$\\[\medskipamount]
Nd$_{2}$CuO$_{4}$ $ab$&&$1248$ &$319$
 (Ref. \onlinecite{hundley89a})&$5900$ (Ref. \onlinecite{fil96a})&$2 \times 1 \times 0.3$&&$5$&$10.7$&$11.0$&$4.9$\\
Nd$_{2}$CuO$_{4}$ $ab$ &(Ref.\,\onlinecite{jin03a}) &&
&&&$275$&$0.27$&$11.1$&$9.6$&$4$\\[\medskipamount]
Sm$_{2}$CuO$_{4}$ $ab$&&$1300$&$353$ (Ref.
\onlinecite{hundley89a})&$5900$&$3.2\times3.7\times0.4 $&
&$1.2$&$16.5$&$21.2$&$5.2$\\[\medskipamount]
Eu$_{2}$CuO$_{4}$ $ab$ &
&$1300$&&$ $&$2.6\times2\times0.3 $&&&&&\\[\medskipamount]
Gd$_{2}$CuO$_{4}$ $ab$ &A&$1292$&$350$&$5900$&$ 2\times0.7\times1.1$&$290$&$11$&$17.5$&$2.4$&$5$\\
Gd$_{2}$CuO$_{4}$ $c$ &A&&&&$ $&&$7.6$&$40.1$&$3.5$&$3.1$\\
Gd$_{2}$CuO$_{4}$ $ab$ &B&&  &&$1.6\times0.9\times0.4$&$295$&$8.3$&$10.6$&$2.6$&$5$\\
\end{tabular}\end{ruledtabular}}
\end{table*}

La$_{2}$CuO$_{4}$ crystallizes in the so-called T structure (also
called K$_2$NiO$_{4}$ structure). The CuO$_4$ plaquettes of the
planes and the apex oxygen ions form CuO$_{6}$ octahedra. At high
temperatures La$_{2}$CuO$_{4}$ is in the high-temperature
tetragonal phase (HTT phase). At $530$\,K a structural phase
transition takes place,\cite{braden94b} where the octahedra tilt
leads to the low-temperature orthorhombic phase (LTO), which is
stable down to lowest temperature. Due to the octahedron tilt the
point bisecting the nearest neighbor Cu-Cu distance  is no longer
a center of inversion symmetry giving rise to a
Dzyaloshinski-Moriya (DM) type interaction.
Sr$_{2}$CuO$_{2}$Cl$_2$ is almost isostructural to La$_2$CuO$_4$
with the La$^{3+}$ ions being substituted by Sr$^{2+}$ and the
apex O$^{2-}$ by Cl$^-$ ions. In Sr$_{2}$CuO$_{2}$Cl$_2$ the HTT
phase is stable down to the lowest temperature and due to
inversion symmetry no DM exchange and consequently no spin
canting occurs. The $R_{2}$CuO$_{4}$ compounds crystallize in the
tetragonal so-called T'-structure. While the T structure may be
viewed as a stacking of one CuO$_2$ layer followed by two LaO (or
SrCl) layers, the stacking of the T' structure is one CuO$_2$
layer followed by a layer of $R^{3+}$ ions, a layer of O$^{2-}$
ions, and finally another layer of $R^{3+}$ ions. Consequently,
there are no apex oxygen ions present in the T' structure and the
basic building blocks are CuO$_4$ plaquettes instead of the
CuO$_{6}$ octahedra of the T structure. For ${\it R} = {\rm Pr}$,
Nd, and Sm the T' structure is stable over the entire temperature
range. For Eu$_{2}$CuO$_{4}$ and Gd$_{2}$CuO$_{4}$ structural
phase transitions are observed with transition temperatures of
$170$\,K and $685$\,K, respectively.\cite{braden94a,vigoureux97b}
The structural changes can be described by an alternating
rotation of the CuO$_4$ plaquettes around the $c$ axis. The
rotation angles amount to $2.3^\circ$ at $20$\,K for $R={\rm Eu}$
and to $5.2\,^\circ$ at $300\,$\,K for $R={\rm
Gd}$.\cite{braden94a,vigoureux97b} The structural transitions
transform both the T and the T' structure into orthorhombic
structures, but all these crystals are usually strongly twinned
with respect to the $a$ and the $b$ axes. As in La$_2$CuO$_4$, the
lower symmetry in the distorted T' phases gives rise to a DM
interaction and leads to a canting of the magnetic moments.

\section{Experimental}
\label{expset}

The $R_2$CuO$_4$ single crystals were grown in Pt crucibles by
the top-seeded solution method as described in
Ref.\,\onlinecite{barilo93a}. The La$_2$CuO$_4$ crystal was grown
by a traveling-solvent floating zone method. The finite DM
interaction for $R={\rm La}$ and Gd causes a weak ferromagnetic
moment, which allows an easy determination of the N\'{e}el
temperature $T_{\rm N}$ by measurements of the magnetic
susceptibility $\chi$. For $R={\rm Gd}$ we find $T_{\rm N} \simeq
292$\,K in agreement with the highest values reported for
Gd$_2$CuO$_4$.\cite{mira95a,klamut94a} Our La$_2$CuO$_{4+\delta}$
single crystal has a comparatively low $T_{\rm N}=245$\,K due to
excess oxygen. A comparison of the in-plane resistivity (not
shown) with Ref.~\onlinecite{yu96a} yields $\delta\approx0.01$.
The determination of $T_{\rm N}$ from $\chi(T)$ does neither work
in the undistorted samples with $R={\rm Pr}$, Nd, and Sm, since
there is no weak ferromagnetism nor for $R={\rm Eu}$, where the
structural transition takes place well below $T_{\rm N}$.
According to Ref.~\onlinecite{jin03a} there is a slope change of
$\chi_{ab}$ at $T_{\rm N}$ in Nd$_2$CuO$_4$, but we could not
reproduce such a feature in $\chi_{ab}$ of our crystal. For
Pr$_2$CuO$_4$ we determined $T_{\rm N} \simeq 250$\,K at the
Laboratoire Leon Brillouin, Saclay by neutron diffraction, which
is well below the maximum values up to $T_{\rm N} \simeq 280$\,K
reported for this compound.\cite{sumarlin95a,matsuda90a,cox89a}
Unfortunately, our Nd$_2$CuO$_4$ crystal is too small for neutron
diffraction and this method cannot be applied for $R= {\rm Eu}$
and Sm because of the large neutron absorption cross section of
these elements.

All crystals have been oriented using a Laue camera and cut into
rectangular pieces. Sample sizes are listed in
Table~\ref{tabelle}. The accuracy of the orientation with respect
to the crystal axes is about~$2^\circ$. Since the
Sm$_{2}$CuO$_{4}$ crystal has approximately the shape of a
cuboid, it has not been cut.  Here, the misalignment amounts to
$\simeq 10^\circ$. The shape of Pr$_{2}$CuO$_{4}$ and
Gd$_{2}$CuO$_{4}$ allowed measurements of $\kappa_{c}$ with a
heat current $j_H$ parallel to the $c$ axis, i.\,e.\
perpendicular to the CuO$_2$ planes, and of $\kappa_{ab}$ with
$j_H$ within the CuO$_2$ planes. For $R = {\rm Nd}$, Sm, and Eu,
we could only measure $\kappa_{ab}$, because these crystals were
very thin with lengths of less $0.4$\,mm  parallel to the $c$
axis. For $R = {\rm Pr}$, Nd, and Gd we measured $\kappa_{ab}$
with $j_H$ parallel to the $a'$ axis of the HTT phase, which has
an angle of 45$^\circ$ with respect to the orthorhombic $a$ and
$b$ axes. For $R = {\rm Eu}$ and Sm $j_H$ had an arbitrary
orientation with respect to the $a$ and $b$ axes.

The thermal conductivity has been measured by a standard
steady-state method. One end of the sample has been attached to
the sample holder by silver paint and a small resistor has been
glued to the opposite end of the sample by an insulating varnish
(VGE-7031, LakeShore). The temperature of the sample holder has
been stabilized and an electrical current through the resistor
has  been used to produce a heat current through the sample. The
resulting temperature gradient has been determined by a
differential Chromel-Au+0.07$\%$Fe-thermocouple, which has been
also glued to the sample. Typical temperature gradients were
about $0.5\%$ of the sample temperature. The absolute accuracy of
our method is restricted to about $10\%$ because of uncertainties
in determining the sample geometry and, in particular, the exact
distance between the two ends of the thermocouple. The relative
accuracy is about one order of magnitude better. Radiation losses
are negligible for our sample geometry.


\section{Results}
\label{results}
\subsection{Gd$_{2}$CuO$_{4}$ and Pr$_{2}$CuO$_{4}$}

\begin{figure}
\begin{center}
\includegraphics[width=0.88\columnwidth,clip]{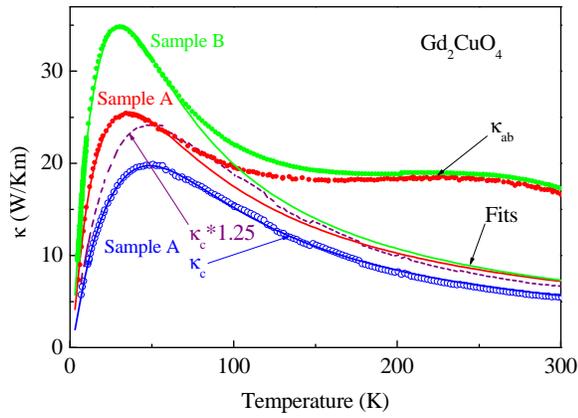}
\end{center}
\caption{(Color online) In-plane ($\kappa_c$) and out-of-plane
($\kappa_{ab}$) thermal conductivity of Gd$_{2}$CuO$_{4}$.
$\kappa_{ab}$ was measured on two different crystals. Solid lines
are fits by the Debye model and the dashed line is $\kappa_c$
multiplied by a factor of $1.25$ (see text).} \label{gd}
\end{figure}

\begin{figure}[b]
\begin{center}
\includegraphics[width=0.88\columnwidth,clip]{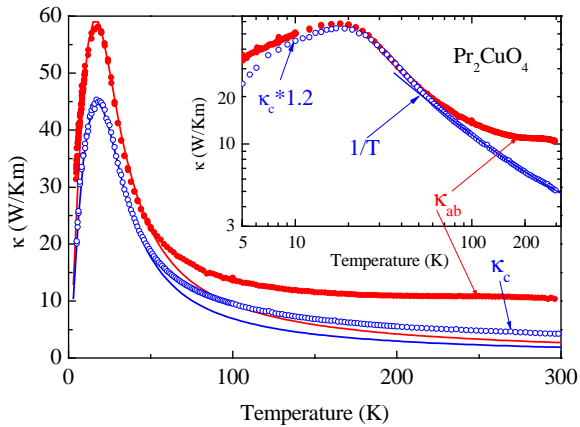}
\end{center}
\caption{(Color online) In-plane ($\kappa_c$) and out-of-plane
($\kappa_{ab}$) thermal conductivity of Pr$_{2}$CuO$_{4}$.  Lines
are fits by the Debye model (see text). Inset: The same data on
double-logarithmic scales with $\kappa_c$ multiplied by a factor
of 1.2. The temperature dependencies of $\kappa_{ab}$ and
$\kappa_c$ around the low-temperature maxima are nearly the same.
Above about 70\,K $\kappa_c$ follows a $1/T$ behavior (solid
line), whereas an anomalous high-temperature upturn is present in
$\kappa_{ab}$.} \label{pr}
\end{figure}

Figure\,\ref{gd} shows the thermal conductivity of
{Gd$_{2}$CuO$_{4}$} measured on two different samples. One sample
allowed to measure $\kappa_{ab}$ and $\kappa_{c}$, whereas the
second crystal was too thin to measure $\kappa_{c}$. In
Figure\,\ref{pr} we display $\kappa_{ab}$ and $\kappa_{c}$ of
{Pr$_{2}$CuO$_{4}$}. Apart from differences around the
low-temperature maximum, which is very sensitive to the sample
quality, our data agree well to the previously reported
$\kappa_{ab}$ of {Pr$_{2}$CuO$_{4}$}.\cite{inyushkin94a} For both
compounds $\kappa_c$ follows the typical temperature dependence
of the thermal conductivity of acoustic phonons. As shown by the
solid lines, $\kappa_c$ can be reasonably well described within a
Debye model. We use the same model and nomenclature as in our
previous publication (see Eqs.~(1) and~(2) of
Ref.~\onlinecite{hofmann03a}) and the corresponding fit
parameters are given in Table~\ref{tabelle}. In both crystals
$\kappa_{ab}$ exceeds $\kappa_{c}$ over the entire temperature
range. The low-temperature maxima for Gd$_2$CuO$_4$ are slightly
shifted in temperature. The most striking anisotropy is, however,
the additional, broad maximum of $\kappa_{ab}$ around 250\,K.
Although the low-temperature maxima of $\kappa_{ab}$ strongly
differ for the two crystals with $R={\rm Gd}$ indicating
differences in the crystal quality, the magnitudes of their
high-temperature maxima are almost identical. In
{Pr$_{2}$CuO$_{4}$} $\kappa_{ab}$ also shows an additional
high-temperature maximum, but its magnitude is less pronounced
(see section \ref{discuss}).

The double-peak structures of $\kappa_{ab}$ cannot be modeled by
the usual Debye model, but it is possible to describe the
low-temperature maxima up to about 50\,K. Comparing the
corresponding fit parameters of $\kappa_{ab}$ and $\kappa_c$ (see
Table~\ref{tabelle}), the largest differences are found for the
parameter $D$, which is significantly larger for the fits of
$\kappa_c$ than for those of $\kappa_{ab}$. This parameter gives
the strength of phonon scattering by planar defects and  it
appears reasonable that due to the layered structure of
$R_2$CuO$_4$ scattering by planar defects should be more
effective for a heat current perpendicular to the planes than for
$j_H$ within the planes. Thus we interpret the different
magnitudes of the low-temperature maxima of $\kappa_{c}$ and
$\kappa_{ab}$ as a consequence of the layered structure. The
high-temperature maxima of $\kappa_{ab}$ around 250\,K will be
discussed in section\,\ref{discuss}.

\subsection{Nd$_{2}$CuO$_{4}$, Sm$_{2}$CuO$_{4}$, and Eu$_{2}$CuO$_{4}$}

\begin{figure}
\begin{center}
\includegraphics[width=0.9\columnwidth,clip]{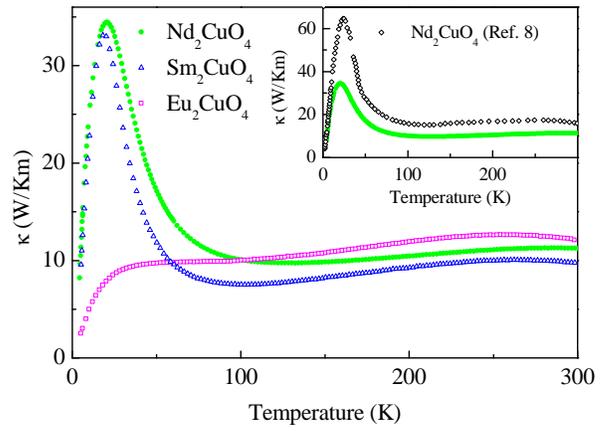} \end{center}
\caption{(Color online) In-plane thermal conductivity of
Nd$_{2}$CuO$_{4}$, Sm$_{2}$CuO$_{4}$, and Eu$_{2}$CuO$_{4}$. For
$R=$ Eu the low-temperature maximum is suppressed.  The inset
compares $\kappa_{ab}$ ($\bullet$) of our Nd$_{2}$CuO$_{4}$
crystal with data  ($\circ$) from Ref.\,\onlinecite{jin03a}.}
\label{nd}
\end{figure}

In Fig.\,\ref{nd} we show the in-plane thermal conductivities
$\kappa_{ab}$ of Nd$_{2}$CuO$_{4}$, Sm$_{2}$CuO$_{4}$, and
Eu$_{2}$CuO$_{4}$. The data for $R={\rm Nd}$ and Sm are very
similar to each other and also to those of $R={\rm Pr}$ and Gd.
In all crystals $\kappa_{ab}$ exhibits a well defined
low-temperature peak around 20\,K and an additional broad maximum
around 250\,K.  For $R={\rm Nd}$ similar results have been
obtained by Jin \textit{et al.}\cite{jin03a}, but $\kappa_{ab}$
of our crystal is systematically lower in the entire temperature
range (see Inset of Fig.\,\ref{nd}). For Eu$_{2}$CuO$_{4}$ the
low-temperature peak of $\kappa_{ab}$ is almost completely
suppressed. As described above, for $R={\rm Eu}$ the structural
transition takes place at 170\,K. The suppression of the
low-temperature peak is most probably a consequence of this
structural instability, which prevents a strong increase of the
phonon mean free path at low temperatures. In contrast, however,
the high-temperature maximum of $\kappa_{ab}$ is hardly affected
by this structural transition. The high-temperature maximum for
Eu$_{2}$CuO$_{4}$ is even more pronounced than for the
structurally stable crystals with $R={\rm Nd}$ and Sm.

\section{Discussion}
\label{discuss}

Our data of $\kappa_c$ of Pr$_{2}$CuO$_{4}$ and Gd$_{2}$CuO$_{4}$
together with $\kappa_c$ of La$_{2}$CuO$_{4}$
(Refs.~\onlinecite{nakamura91a,yan03a,sun03a,hess03a}) clearly
reveal that, on the one hand, the out-of-plane thermal
conductivities of all these compounds of slightly different
structures (tetragonal and orthorhombic T', and orthorhombic T)
do not exhibit any indications of an anomalous high-temperature
contribution. On the other hand, all in-plane conductivities
clearly show additional broad high-temperature maxima. The fact
that these high-temperature maxima are present in crystals without
($R_{2}$CuO$_{4}$ with $R={\rm Pr}$, Nd, Sm, and
Sr$_{2}$CuO$_{2}$Cl$_{2}$) or with (different) structural
instabilities (T: $R={\rm La}$; T': $R={\rm Eu}$ and Gd)
unambiguously shows that the anomalous contribution to
$\kappa_{ab}$ does not depend on the existence of structural
instabilities. This complements our previous suggestion based on
a comparative study of $\kappa $ of Sr$_{2}$CuO$_{2}$Cl$_{2}$ and
La$_{2}$CuO$_{4}$ and, as we have discussed in detail in
Ref.~\onlinecite{hofmann03a}, the most natural explanation for
the high-temperature maximum of $\kappa_{ab}$ is an additional
contribution to the heat transport caused by magnetic excitations.

The similar behavior of $\kappa$ for all $R_{2}$CuO$_{4}$
confirms that such a magnetic contribution $\kappa_{mag}$ to the
in-plane heat transport is indeed an intrinsic property of the
CuO$_2$ planes. It remains, however, to clarify what determines
the magnitude of $\kappa_{mag}$. For a quantitative analysis we
consider the in-plane thermal conductivity as the sum of a
phononic and a magnetic contribution
\begin{equation}
\kappa_{ab}=\kappa_{ph}+\kappa_{mag},
\label{kmag}
\end{equation}
which are only weakly coupled to each other. In general, such an
Ansatz can be used when the characteristic energy scales for the
two contributions are well separated from each other. For
example, this is usually the case for electronic and phononic
heat transport since the Fermi temperature is much larger than the
Debye temperature, i.e.\ $T_F\gg \Theta_D$. In the case of
$R_2$CuO$_4$ it is a priori not clear whether the assumption of
weakly coupled phononic and magnetic contributions is fulfilled,
since the magnetic coupling $J$ is only about four times as large
as $\Theta_D$ (see Table\,\ref{tabelle}). However, the
experimental observation that $\kappa_{ab}$ shows two
characteristic maxima, which are well separated from each other,
encourages us to use Eq.\,(\ref{kmag}).

In order to separate $\kappa_{\rm mag}$ from $\kappa_{ab}$ we
assume that in the region of the low-temperature peak
$\kappa_{\rm mag}$ is negligibly small and fit the data for
$T<50$\,K (Gd$_{2}$CuO$_{4}$: $T<85$\,K) by the Debye model (see
Eq.~(1) of Ref.~\onlinecite{hofmann03a}) and subtract the
extrapolation of the fit from the measured data up to room
temperature, i.\,e.\ $\kappa_{\rm mag}=\kappa_{ab}-\kappa_{\rm
ph}^{fit}$.\cite{params} This analysis is not possible for
Eu$_{2}$CuO$_{4}$, because the low-temperature maximum is not
well-enough pronounced. However, even without a fit we expect
rather similar values of $\kappa_{\rm mag}$ for $R={\rm Nd}$, Sm,
and Eu, because the high-temperature data of $\kappa_{ab}$ are
very similar for these crystals (see Fig.\,\ref{nd}). One can
check the applicability of the Debye model to describe the
phononic contribution by corresponding fits of $\kappa_c$, i.e.\
we restrict the temperature range of the fit to the
low-temperature maxima and then compare the high-temperature
extrapolations of the fits to the measured $\kappa_c$. As shown
in Figs.\,\ref{gd} and~\ref{pr} these fits yield a good
description of $\kappa_c$ for $R={\rm Gd}$ over the entire
temperature range, whereas in the case of Pr the high-temperature
values of $\kappa_c$ are slightly underestimated by the fit. This
probably arises from the sharper low-temperature peak for $R={\rm
Pr}$. Since the low-temperature peaks of $\kappa_{ab}$ for $R={\rm
Pr}$, Nd, and Sm are also rather sharp, one may expect that the
corresponding Debye fits will also underestimate the
high-temperature values of the phononic contribution of
$\kappa_{ab}$ and consequently the magnetic contributions
$\kappa_{\rm mag}$ may be overestimated to some extent.

\begin{figure}
\begin{center}
\includegraphics[width=0.9\columnwidth,clip]{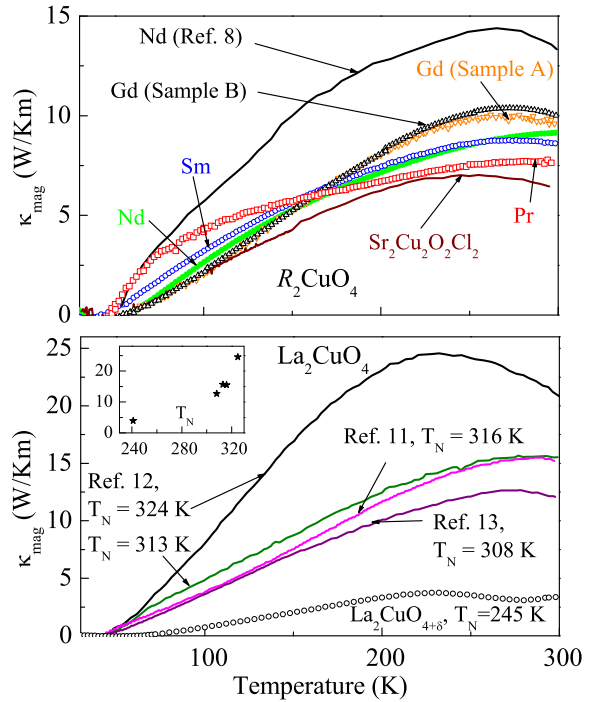} \end{center}
\caption{(Color online) Magnetic contributions to the in-plane
thermal conductivity, calculated via
$\kappa_{mag}=\kappa_{ab}-\kappa_{ph}$, where $\kappa_{ph}$ is
determined by a Debye fit of the low-temperature maximum. Upper
panel: Values calculated from our measurements of $\kappa_{ab}$ of
$R_2$CuO$_4$ and from the data from Ref. \onlinecite{jin03a}.
Lower panel: The same analysis for various data of
La$_2$CuO$_{4+\delta}$ taken from
Refs.\,\onlinecite{nakamura91a,yan03a,sun03a} and of our crystal
with $T_{\rm N}=245$\,K. Inset: The maximum of the calculated
$\kappa_{\rm mag}$ $vs.$ the N\'{e}el temperature for
La$_2$CuO$_{4+\delta}$ (see text).} \label{mag}
\end{figure}

In the upper panel of Fig.\,\ref{mag} we compare the resulting
$\kappa_{\rm mag}$ of all our $R_2$CuO$_4$ crystals, our previous
result of Sr$_2$CuO$_2$Cl$_2$ (Ref.\,\onlinecite{hofmann03a}), and
$\kappa_{\rm mag}$ obtained from an analysis of $\kappa_{ab}$
measured on Nd$_2$CuO$_4$ (Ref.\,\onlinecite{jin03a}). For
comparison, we also show $\kappa_{\rm mag}$ obtained from the
literature data of various La$_2$CuO$_4$
crystals.\cite{nakamura91a,yan03a,sun03a} Obviously, the
temperature dependence of $\kappa_{\rm mag}$ is very similar for
all crystals, but the magnitude of the broad maximum varies
between about 7 and 10\,W/Km for our Sr$_2$CuO$_2$Cl$_2$ and
$R_2$CuO$_4$ crystals and from about 12 to 25\,W/Km for the
various crystals from literature with $R={\rm La}$ and
Nd.\cite{jin03a,nakamura91a,yan03a,sun03a} Although these
differences are not too large, we do not think that they simply
arise from the experimental uncertainty in the quantitative
determination of $\kappa_{\rm mag}$.

Because the magnetic properties are rather similar for the
different crystals, we expect that the different $\kappa_{mag}$
mainly arise from differences in the scattering of the magnetic
excitations. Possible scattering mechanisms are scattering
between magnetic excitations and scattering by defects, phonons,
and charge carriers. One may suspect that scattering between
magnetic excitations, comparable e.\,g.\ to phonon-phonon Umklapp
scattering, plays the most important role with respect to the
temperature dependence of $\kappa_{mag}$. A deeper analysis of
this scattering requires a detailed theoretical model for the
dynamics of magnetic excitations, but even without such a model
one may conclude that the similar magnetic properties naturally
explain the similar temperature dependencies of $\kappa_{mag}$ of
the different compounds.

The influence of defects and charge carriers on the thermal
conductivity has been investigated in Zn- and Sr-doped
La$_2$CuO$_4$.\cite{sun03a,hess03a}. It has been found that Sr
doping suppresses $k_{\rm mag}$ much stronger than Zn doping and
$k_{mag}$ vanishes almost completely above about $1$\% Sr. For
both dopings the magnetic system is diluted, either by replacing
magnetic Cu$^{2+}$ by nonmagnetic Zn$^{2+}$ ions or by the
formation of Zhang-Rice singlets due to the introduced holes.
However, the mobility of the holes strongly enhances the effect of
charge-carrier doping, what is also reflected in a much stronger
suppression of $T_{\rm N}$ by Sr doping as compared to Zn
doping.\cite{hucker99a} As shown in the lower panel of
Fig.\,\ref{mag}, the magnitude of $\kappa_{\rm mag}$ for the
various nominally undoped La$_2$CuO$_{4}$ crystals varies by
about a factor of two. Since it is known that
La$_2$CuO$_{4+\delta}$ is likely to have some excess oxygen, we
plot the maximum of $\kappa_{\rm mag}$ as a function of $T_{\rm
N}$, which is very sensitive to small amounts of $\delta $ (Inset
of Fig.\,\ref{mag}). The observed correlation between the
magnitude of $\kappa_{\rm mag}$ and $T_{\rm N}$ is a clear
indication that the different magnitudes of $\kappa_{\rm mag}$
arise from small amounts of charge carriers in the different
crystals.\cite{yan03a,oxygenorder}

One may suspect that the higher values of $\kappa_{\rm mag}$ of
$R_2$CuO$_{4}$ with $R= {\rm La}$ and Nd from
Refs.~\onlinecite{jin03a,nakamura91a,yan03a,sun03a} in comparison
to our crystals could result from a weak charge carrier doping in
our crystals. The rather low $T_{\rm N} \simeq 250$\,K of our
Pr$_2$CuO$_4$ crystal compared to the $T_{\rm N}$ values up to
$\simeq 280$\,K reported in
literature\cite{sumarlin95a,matsuda90a,cox89a} supports this
view. However, this argumentation can neither explain the low
$\kappa_{\rm mag}$ of our Gd$_2$CuO$_4$ with a large $T_{\rm N}
\simeq 292$\,K nor does it hold for Sr$_2$CuO$_2$Cl$_2$, which is
commonly believed to be very stable with respect to charge
carrier doping. Unfortunately, not much is known about possible
variations of the oxygen content in $R_2$CuO$_4$. Irrespective of
the question of the exact oxygen stoichiometry, our finding that
$\kappa_{\rm mag}$ is very similar in crystals with and without
structural instabilities leads to the conclusion that scattering
by phonons seems to play a minor role for the magnetic heat
transport in the CuO$_2$ planes. This is most clearly seen in
Eu$_2$CuO$_4$ where the phononic low-temperature peak of
Eu$_2$CuO$_4$ is strongly suppressed by a structural instability
whereas its magnetic high-temperature maximum is hardly affected.

\section{conclusions}

In summary, we have studied the thermal conductivity of the rare
earth cuprates $R_{2}$CuO$_{4}$ for both, a heat current
perpendicular ($R={\rm Pr}$ and Gd) and parallel ($R={\rm Pr}$,
Nd, Sm, Eu, and Gd) to the CuO$_2$ planes. The out-of-plane
thermal conductivity shows the typical temperature dependence of
a purely phononic thermal conductivity with a low-temperature
maximum, whose magnitude depends on the crystal quality. In
contrast, the in-plane conductivity for all crystal exhibits a
pronounced double-peak structure consisting (i) of a
low-temperature peak similar to that of the out-of-plane thermal
conductivity and (ii) of an anomalous high-temperature
contribution with a broad maximum around 250\,K. Such an
anisotropy between the in-plane and the out-of-plane thermal
conductivity is also found in
La$_{2}$CuO$_{4}$.\cite{nakamura91a,yan03a,sun03a,hess03a} The
fact that the double-peak is present in the structurally stable
$R_{2}$CuO$_{4}$ with $R={\rm Pr}$, Nd, and Sm unambiguously
rules out the possibility that the double-peak structure is
caused by a structural instability, which is present for $R={\rm
La}$, Eu, and Gd. The qualitative anisotropy between the in-plane
and the out-of-plane thermal conductivity and the rather similar
high-temperature behavior of the out-of-plane thermal
conductivity for all the different crystals gives clear evidence
that this additional high-temperature contribution arises from a
sizeable heat transport by magnetic excitations within the
CuO$_2$ planes. Our analysis yields a magnetic contribution to
the in-plane thermal conductivity  between about 7 to 25\,W/Km
depending on the $R$ system. In weakly doped La$_{2}$CuO$_{4}$
this magnetic contribution is strongly suppressed showing that
scattering of magnetic excitations by mobile charge carriers
plays an important role. In contrast, the structural instability
does hardly influence the magnetic thermal conductivity
indicating that scattering of magnetic excitations by soft or
anharmonic phonons plays a minor role. In order to clarify the
role of scattering between magnetic excitations theoretical
models describing the dynamics of magnetic excitations would be
highly desirable.

\begin{acknowledgments}
We acknowledge useful discussions with M.~Braden, M.~Gr\"uninger,
and A.~Sologubenko. This work was supported by the Deutsche
Forschungsgemeinschaft through SFB 608. The work in Minsk was
supported in part by the Belarussian Foundation for Fundamental
Research by grant No. BRFFI  F05-129.
\end{acknowledgments}


\begin{thebibliography}{10}
\bibitem{roomtemp}
Here, only room temperature values are listed because temperature
dependent
  values are not available for all samples. Note that these values are slightly
  larger then the low-temperature data used in Ref. \onlinecite{hofmann03a},
  which were obtained by bi-magnon-plus-phonon absorption from infrared
  spectroscopy for La$_{2}$CuO$_{4}$ and Sr$_{2}$CuO$_{2}$Cl$_2$.
\bibitem{params}
We do not treat $\theta_{D}$, $v_s$ and $L$ as free fit
parameters, but use
  experimentally determined values. If no experimental values are available as
  e.\,g.\ for Gd$_2$CuO$_4$ we took values similar to those obtained in the
  other compounds (see Table\,\ref{tabelle}). The resulting fits hardly change
  when the upper boundary of the temperature range is varied by $\pm 10$\,K.
\bibitem{oxygenorder}
For stronger oxygen doping a simple determination of $\delta $
from the
  measured $T_{\rm N}$ is not possible, because for $\delta\gtrsim 0.01$ a
  phase separation into oxygen-rich and oxygen-poor regions occurs, which is
  accompanied by various phase transitions leading to a complex phase
  diagram.\cite{yu96a} The kink in $\kappa_{ab}$ at $T\approx 290$\,K in our
  La$_2$CuO$_{4+\delta}$ crystal is also related to this phase separation.
\end{thebibliography}

\end{document}